\documentclass[letterpaper]{article}
\usepackage{aaai}
\usepackage{times}
\usepackage{helvet}
\usepackage{courier}
\usepackage{subfig}
\usepackage{graphicx}
\graphicspath{{Figures/}}

\title{The Pulse of News in Social Media: Forecasting Popularity}
\author{Roja Bandari\\
Department of Electrical Engineering\\ UCLA \\ Los Angeles, CA 90024\\roja@ucla.edu
\And Sitaram Asur\\
Social Computing Lab\\HP Labs\\ Palo Alto, CA 94304 \\ sitaram.asur@hp.com
\And Bernardo A. Huberman\\ 
Social Computing Lab\\HP Labs\\ Palo Alto, CA 94304 \\bernardo.huberman@hp.com
}

\begin{document}
\maketitle

\begin{abstract}
\begin{quote}

News articles are extremely time sensitive by nature. There is also intense competition among news items to propagate as widely as possible. Hence, the task of predicting the popularity of news items on the social web is both interesting and challenging. Prior research has dealt with predicting eventual online popularity based on early popularity. It is most desirable, however, to predict the popularity of items prior to their release, fostering the possibility of appropriate decision making to modify an article and the manner of its publication. In this paper, we construct a multi-dimensional feature space derived from properties of an article and evaluate the efficacy of these features to serve as predictors of online popularity. We examine both regression and classification algorithms and demonstrate that despite randomness in human behavior, it is possible to predict ranges of popularity on twitter with an overall 84\% accuracy. Our study also serves to illustrate the differences between traditionally prominent sources and those immensely popular on the social web.

\end{quote}
\end{abstract}

\section{Introduction}

News articles are very dynamic due to their relation to continuously developing events that typically have short lifespans. For a news article to be popular, it is essential for it to propagate to a large number of readers within a short time. Hence there exists a competition among different sources to generate content which is relevant to a large subset of the population and becomes virally popular.

Traditionally, news reporting and broadcasting has been costly, which meant that large news agencies dominated the competition. But the ease and low cost of online content creation and sharing has recently changed the traditional rules of competition for public attention. News sources now concentrate a large portion of their attention on online mediums where they can disseminate their news effectively and to a large population. It is  therefore common for almost all major news sources to have active accounts in social media services like Twitter to take advantage of the enormous reach these services provide.

Due to the time-sensitive aspect and the intense competition for attention, accurately estimating the extent to which a news article will spread on the web is extremely valuable to journalists, content providers, advertisers, and news recommendation systems. This is also important for activists and politicians who are using the web increasingly more to influence public opinion. 

However, predicting online popularity of news articles is a challenging task. First, \emph{context} outside the web is often not readily accessible and elements such as local and geographical conditions and various circumstances that affect the population make this prediction difficult.  Furthermore, \emph{network properties} such as the structure of social networks that are propagating the news, influence variations among members, and interplay between different sections of the web add other layers of complexity to this problem. Most significantly, intuition suggests that the \emph{content} of an article must play a crucial role in its popularity. Content that resonates with a majority of the readers such as a major world-wide event can be expected to garner wide attention while specific content relevant only to a few may not be as successful.

Given the complexity of the problem due to the above mentioned factors, a growing number of recent studies \cite{DBLP:journals/cacm/SzaboH10}, \cite{DBLP:conf/webi/LeeMS10}, \cite{Tatar2011}, \cite{6036808},  \cite{DBLP:conf/www/LermanH10} make use of early measurements of an item's popularity to predict its future success. In the present work we investigate a more difficult problem, which is prediction of social popularity without using early popularity measurements, by instead solely considering features of a news article \emph{prior} to its publication. We focus this work on observable features in the content of an article as well as its source of publication. Our goal is to discover if any predictors relevant only to the content exist and if it is possible to make a reasonable forecast of the spread of an article based on content features.

The news data for our study was collected from Feedzilla~\footnote{www.feedzilla.com} --a news feed aggregator-- and measurements of the spread are performed on Twitter~\footnote{www.twitter.com}, an immensely popular microblogging social network. Social popularity for the news articles are measured as the number of times a news URL is posted and shared on Twitter. 

To generate features for the articles, we consider four different characteristics of a given article. Namely: 
\begin{itemize}
\item The news source that generates and posts the article
\item The category of news this article falls under
\item The subjectivity of the language in the article
\item Named entities mentioned in the article
\end{itemize}
We quantify each of these characteristics by a score making use of different scoring functions. We then use these scores to generate predictions of the spread of the news articles using regression and classification methods. Our experiments show that it is possible to estimate ranges of popularity with an overall accuracy of 84\% considering only content features. Additionally, by comparing with an independent rating of news sources, we demonstrate that there exists a sharp contrast between traditionally popular news sources and the top news propagators on the social web.

In the next section we provide a survey of recent literature related to this work. Section 3 describes the dataset characteristics and the process of feature score assignment. In Section 4 we will present the results of prediction methods. Finally, in Section 5 we will conclude the paper and discuss future possibilities for this research.

\section{Related Work}
 
Stochastic models of information diffusion as well as deterministic epidemic models have been studied extensively in an array of papers, reaffirming theories developed in sociology such as diffusion of innovations \cite{rogers1995diffusion}. Among these are models of viral marketing \cite{DBLP:journals/tweb/LeskovecAH07}, models of attention on the web \cite{Wu06112007}, cascading behavior in propagation of information \cite{DBLP:journals/sigkdd/GruhlLGT04} \cite{Leskovec07cascadingbehavior} and models that describe heavy tails in human dynamics \cite{PhysRevE.73.036127}. While some studies incorporate factors for content \emph{fitness} into their model \cite{Simkin2008}, they only capture this in general terms and do not include detailed consideration of content features.

\citeauthor{salganik2010} performed a controlled experiment on music, comparing quality of songs versus the effects of social influence\cite{salganik2010}. They found that song quality did not play a role in popularity of highly rated songs and it was social influence that shaped the outcome. The effect of user influence on information diffusion motivates another set of investigations \cite{DBLP:conf/kdd/KempeKT03}, \cite{ICWSM101530},\cite{Agarwal:2008:IIB:1341531.1341559}, \cite{DBLP:conf/www/LermanH10}.

On the subject of news dissemination, \cite{DBLP:conf/kdd/LeskovecBK09} and \cite{DBLP:conf/wsdm/YangL11} study temporal aspects of spread of news memes online, with \cite{DBLP:conf/icwsm/LermanG10} empirically studying spread of news on the social networks of digg and twitter and \cite{DBLP:conf/icwsm/SunRML09} studying facebook news feeds.

A growing number of recent studies predict spread of information based on early measurements (using early votes on digg, likes on facebook, click-throughs, and comments on forums and sites). \cite{DBLP:journals/cacm/SzaboH10} found that eventual popularity of items posted on youtube and digg has a strong correlation with their early popularity;  \cite{DBLP:conf/webi/LeeMS10} and \cite{Tatar2011} predict the popularity of a discussion thread using features based on early measurements of user comments. \cite{6036808} propose the notion of a virtual temperature of weblogs using early measurements. \cite{DBLP:conf/www/LermanH10} predict digg counts using stochastic models that combine design elements  of the site -that in turn lead to collective user behavior- with information from early votes.

Finally, recent work on variation in the spread of content has been carried out by \cite{Romero2011} with a focus on categories of twitter hashtags (similar to keywords).  This work is aligned with ours in its attention to importance of content in variations among popularity, however they consider categories only, with news being one of the hashtag categories.  \cite{DBLP:conf/sbp/YuCK11} conduct similar work on social marketing messages.

\section{Data and Features}

This section describes the data, the feature space, and feature score assignment in detail. 

\subsection{Dataset Description}
\label{sec:Data}
Data was collected in two steps: first, a set of articles were collected via a news feed aggregator, then the number of times each article was linked to on twitter was found. In addition, for some of the feature scores, we used a 50-day history of posts on twitter. The latter will be explained in section \ref{sec:Feature Description and Scoring} on feature scoring.

\begin{figure}[h] 
   \centering 
   \includegraphics[width=2.5in, trim= 0cm 1cm 0cm 2cm]{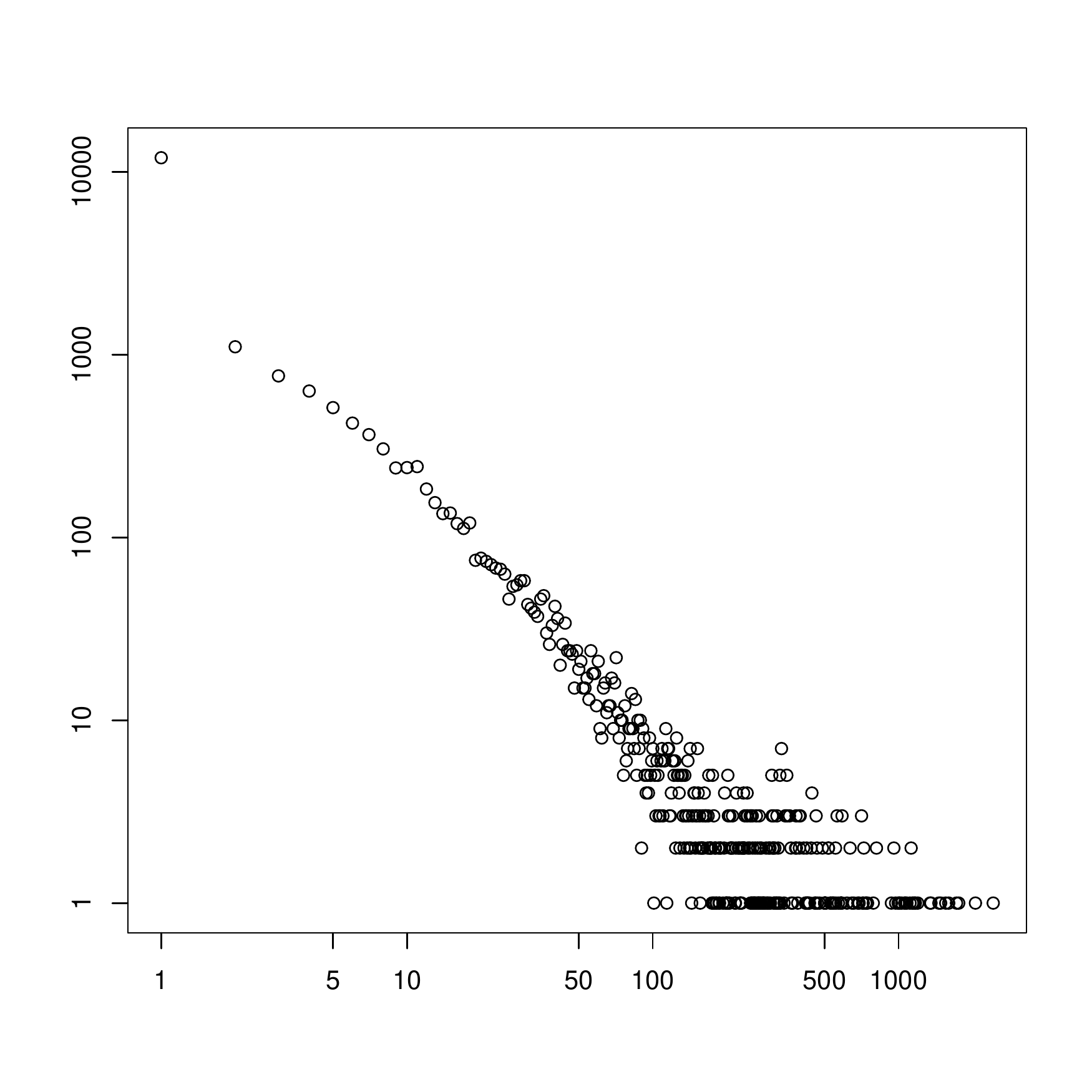} 
   \caption{Log-log distribution of tweets.} 
   \label{fig:powerLaw100}
\end{figure}

Online news feed aggregators are services that collect and deliver news articles as they are published online. Using the API for a news feed aggregator named Feedzilla, we collected news feeds belonging to all news articles  published online during one week (August 8th to 16th, 2011). The feed for an article includes a title, a short summary of the article, its url, and a time-stamp. In addition, each article is pre-tagged with a category either provided by the publisher or in some manner determined by Feedzilla. 
A fair amount of cleaning was performed to remove redundancies, resolve naming variations, and eliminate spam through the use of automated methods as well as manual inspection. As a result over 2000 out of a total of 44,000 items in the data were discarded. 
\begin{figure*}[ht] 
   \centering
   \includegraphics[width=7in, trim = 0 2cm 0cm 0cm]{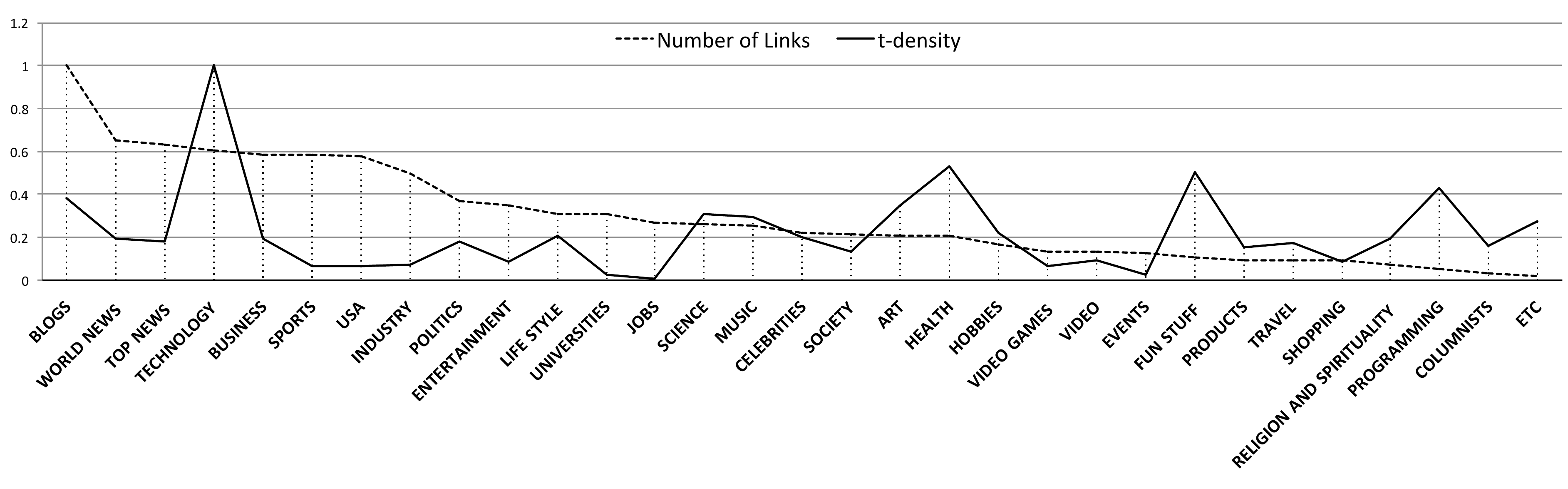} 
   \caption{Normalized t-density scores for categories}
   \label{fig:categories}
\end{figure*}

The next phase of data collection was performed using Topsy \footnote{http://topsy.com} , a Twitter search engine that searches all messages posted on Twitter. We queried for the number of times each news link was posted or reshared on Twitter (tweeted or retweeted). Earlier research \cite{DBLP:conf/kdd/LeskovecBK09} on news meme buildup and decay suggest that popular news threads take about 4 days until their popularity starts to plateau. Therefore, we allowed 4 days for each link to fully propagate before querying for the number of times it has been shared. 
 
The first half of the data was used in category score assignment (explained in the next section). The rest we partitioned equally into 10,000 samples each for training and test data for the classification and regression algorithms. Figure \ref{fig:powerLaw100} shows the log distribution of total tweets over all data, demonstrating a long tail shape which is in agreement with other findings on distribution of Twitter information cascades \cite{Zhou:2010:IRT:1964858.1964875}. The graph also shows that articles with zero tweets lie outside of the general linear trend of the graph because they did not propagate on the Twitter social network. 

Our objective is to design features based on content to predict the number of tweets for a given article. In the next section we will describe these features and the methods used to assign values or scores to features.

\subsection{Feature Description and Scoring}
\label{sec:Feature Description and Scoring}

Choice of features is motivated by the following questions: Does the category of news affect its popularity within a social network? Do readers prefer factual statements or do they favor personal tone and emotionally charged language? Does it make a difference whether famous names are mentioned in the article? Does it make a difference who publishes a news article?

These questions motivate the choice of the following characteristics of an article as the feature space: the category that the news belongs to (e.g. politics, sports, etc.), whether the language of the text is objective or subjective, whether (and what) named entities are mentioned, and what is the source that published the news. These four features are chosen based on their availability and relevance, and although it is possible to add any other available features in a similar manner, we believe the four features chosen in this paper to be the most relevant. 

We would like to point out that we use the terms article and link interchangeably since each article is represented by its URL link.

\subsubsection{Category Score}
News feeds provided by Feedzilla are pre-tagged with category labels describing the content. We adopted these category labels and designed a score for them which essentially represents a prior disribution on the popularity of categories.
Figure \ref{fig:categories} shows a plot of categories and the number of article links in each category. We observe that news related to Technology has a more prominent presence in our dataset and most probably on twitter as a whole. Furthermore, we can see categories (such as Health) with low number of published links but higher rates of tweet per link. These categories perhaps have a niche following and loyal readers who are intent on posting and retweeting its links.

Observing the variations in average tweets per link from Figure~\ref{fig:categories} we use this quantity to represent the prior popularity for a category. In order to assign a value (i.e. score) to each category, we use the the first 22,000 points in the dataset to compute the average tweet per article link in that category. We call this average tweet per link the \emph{t-density} score and we will use this measure in score assignments for some other features as well. 

\subsubsection{Subjectivity}

\begin{figure}[h]
\begin{center}
\includegraphics[width=2.5in, trim = 0cm 1cm 0cm 1cm]{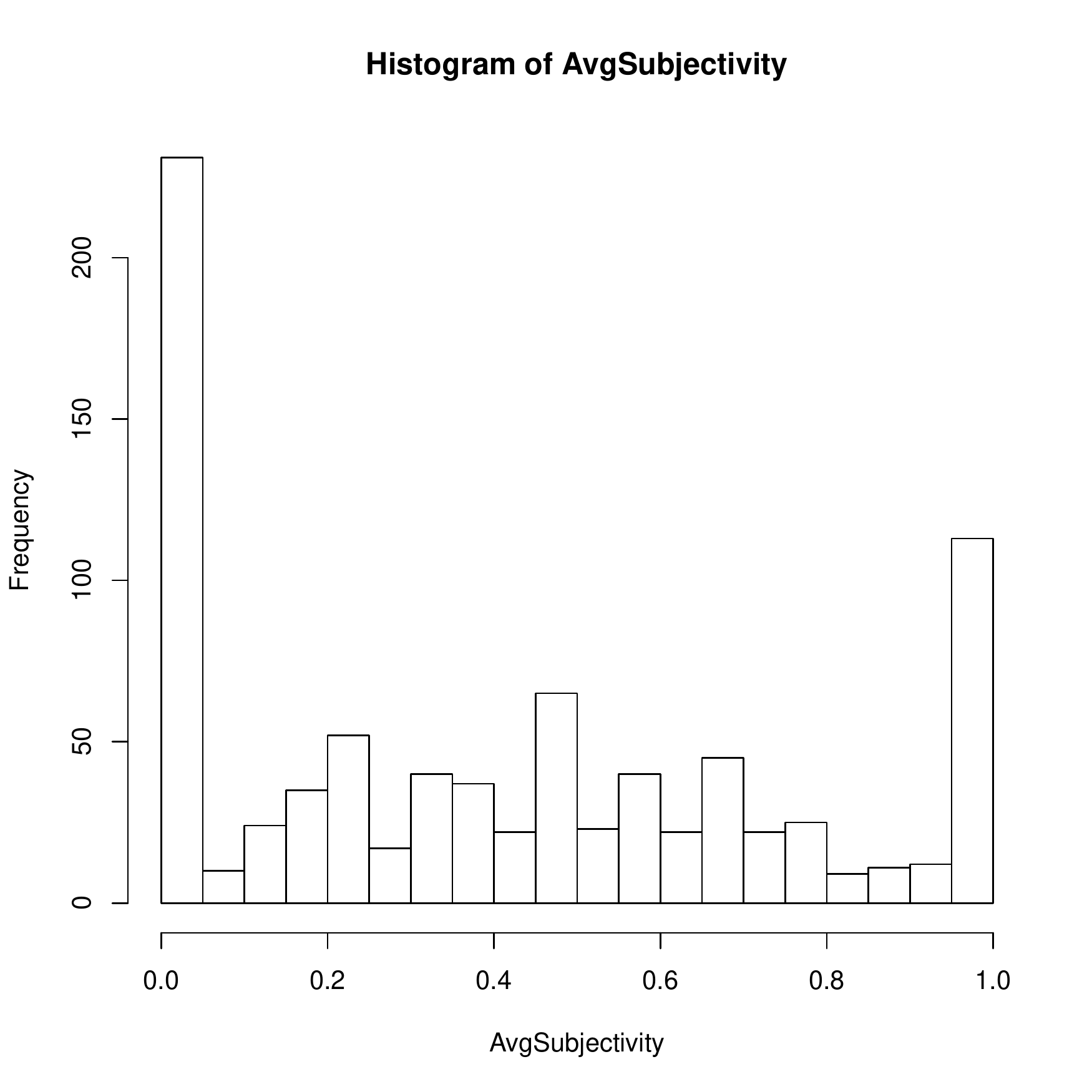} 
\end{center}
\caption{Distribution of average subjectivity of sources.}
\label{fig:subjectivity}
\end{figure}

Another feature of an article that can affect the amount of online sharing is its language. We want to examine if an article 
written in a more emotional, more personal, and more subjective voice can resonate stronger with the readers. Accordingly, we design a binary feature for subjectivity where we assign a zero or one value based on whether the news article or commentary is  written in a more subjective voice, rather than using factual and objective language. We make use of a subjectivity classifier from LingPipe~\cite{lingpipe} a natural language toolkit. Since this requires training data, we use transcripts from well-known tv and radio shows belonging to Rush Limbaugh \footnote{http://www.rushlimbaugh.com} and Keith Olberman \footnote{http://www.msnbc.msn.com/id/32390086} as the corpus for subjective language. On the other hand, transcripts from CSPAN \footnote{http://www.c-span.org} as well as the parsed text of a number of articles from the website FirstMonday \footnote{http://firstmonday.org} are used as the training corpus for objective language. The above two training sets provide a  very high training accuracy of 99\% and manual inspection of final results confirmed that the classification was satisfactory.  Figure \ref{fig:subjectivity} illustrates the distribution of average subjectivity per source, showing that some sources consistently publish news in a more objective language and a somwhat lower number in a more subjective language. 
 
\subsubsection{Named Entities}
In this paper, a named entity refers to a known place, person, or organization. Intuition suggests that mentioning well-known entities can affect the spread of an article, increasing its chances of success. For instance, one might expect articles on Obama to achieve a larger spread than those on a minor celebrity. And it has been well documented that fans are likely to share almost any content on celebrities like Justin Bieber, Oprah Winfrey or Ashton Kutcher. 
We made use of the Stanford-NER ~\footnote{http://nlp.stanford.edu/software/CRF-NER.shtml} entity extraction tool to extract all the named entities present in the title and summary of each article. We then assign scores to over 40,000 named entities by studying historical prominence of each entity on twitter over the timeframe of a month. The assigned score is the average t-density (tweet per link) of each named entity. To assign a score for a given article we use three different values: the number of named entities in an article, the highest score among all the named entities in an article, and the average score among the entities. 

\subsubsection{Source Score}

The data includes articles from 1350 unique sources on the web. We assign scores to each source based on the historical success of each source on Twitter. For this purpose, we collected the number of times articles from each source were shared on Twitter in the past. We used two different scores, first the aggregate number of times articles from a source were shared, and second the t-density of each source (average number of times each article belonging to a source was shared). The latter proved to be a better score assignment compared to the aggregate.  

\begin{figure}[h]
\begin{center}
\includegraphics[width=2.5in, trim = 0cm 1cm 0cm 1cm]{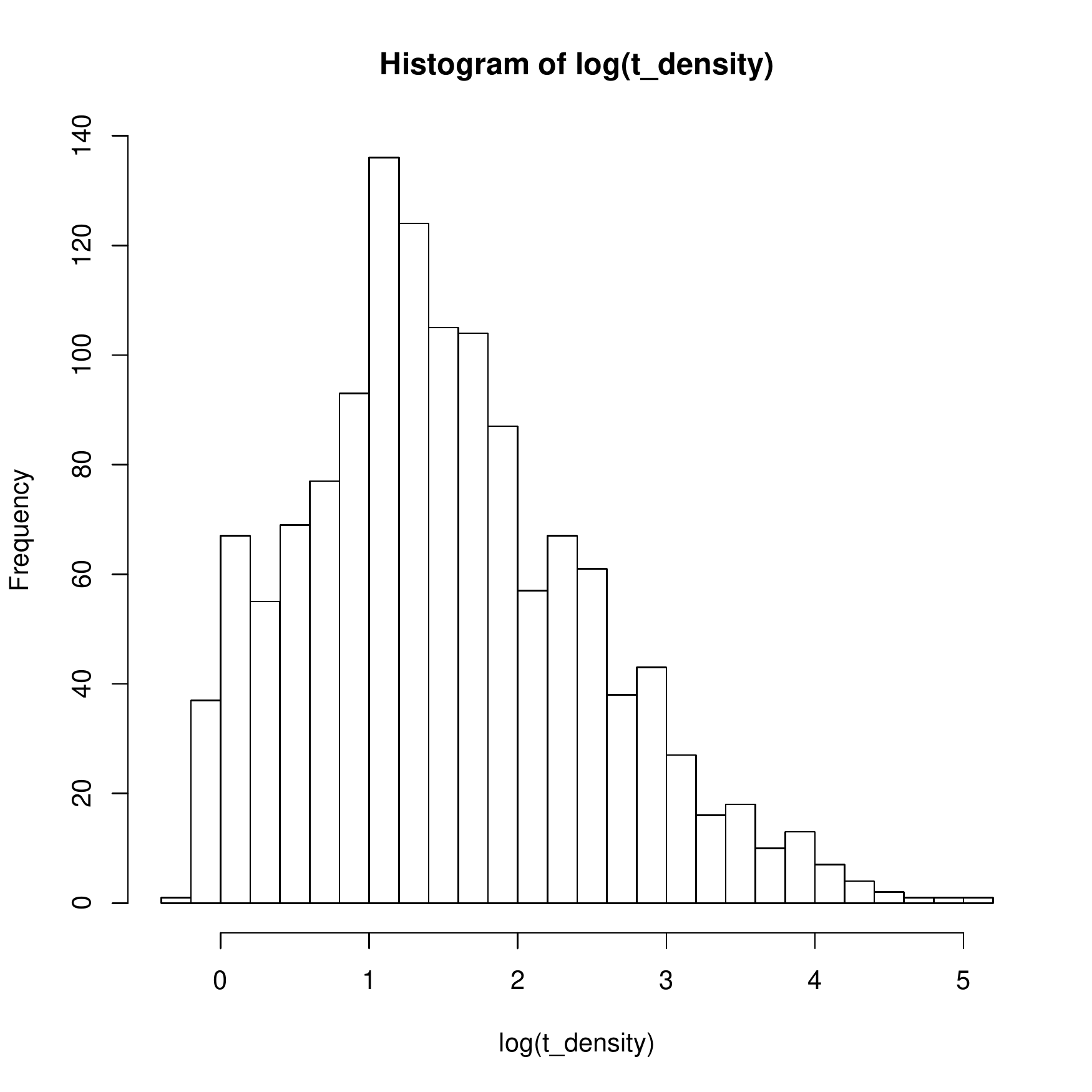}
\end{center}
\caption{Distribution of log of source t-density scores}
\label{fig:sourceScoredist}
\end{figure}

\begin{figure}[h] 
   \centering 
   \includegraphics[width=3in, trim = 1cm 0 0cm 0]{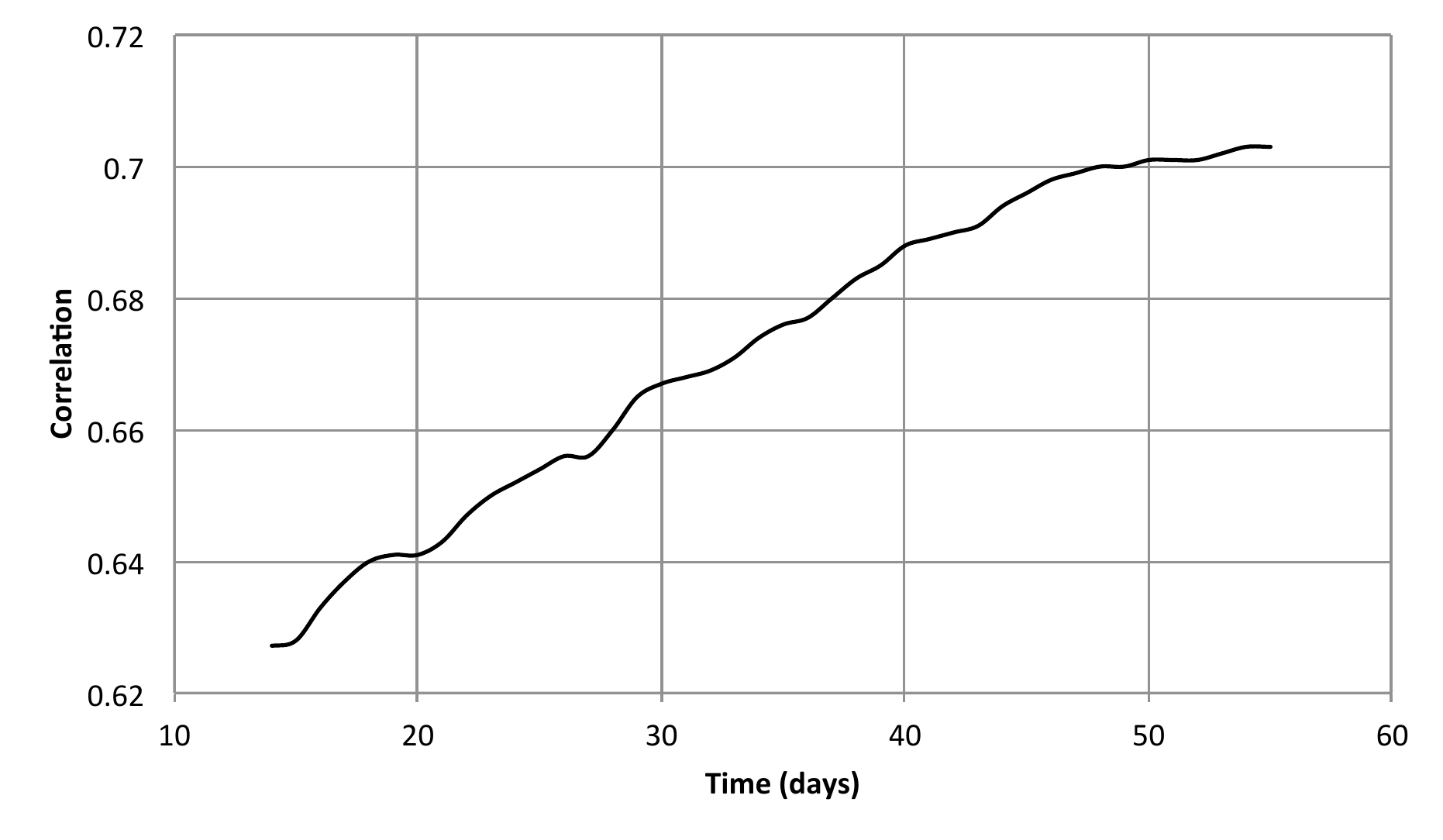} 
   \caption{Correlation trend of source scores with t-density in data. Correlation increases with more days of historical data until it plateaus after 50 days.}
   \label{fig:sourceCor}
\end{figure}

To investigate whether it is better to use a smaller portion of more recent history, or a larger portion going back farther in time and possibly collecting outdated information, we start with the two most recent weeks prior to our data collection and increase the number of days, going back in time.  Figure \ref{fig:sourceCor} shows the trend of correlation between the t-density of sources in historical data and their true t-density of our dataset. We observe that the correlation increases with more datapoints from the history until it begins to plateau near 50 days. Using this result, we take 54 days of history prior to the first date in our dataset. We find that the correlation of the assigned score found in the above manner has a correlation of 0.7 with the t-density of the dataset. Meanwhile, the correlation between the source score and number of tweets of any given article is 0.35, suggesting that information about the source of publication alone is not sufficient in predicting popularity. Figure \ref{fig:sourceScoredist} shows the distribution of log of source scores (t-density). Taking the log of source scores produces a more normal shape, leading to improvements in regression algorithms.

\begin{figure}[h] 
   \centering
   \includegraphics[width=3.2in , trim = 1cm 0 0cm 0]{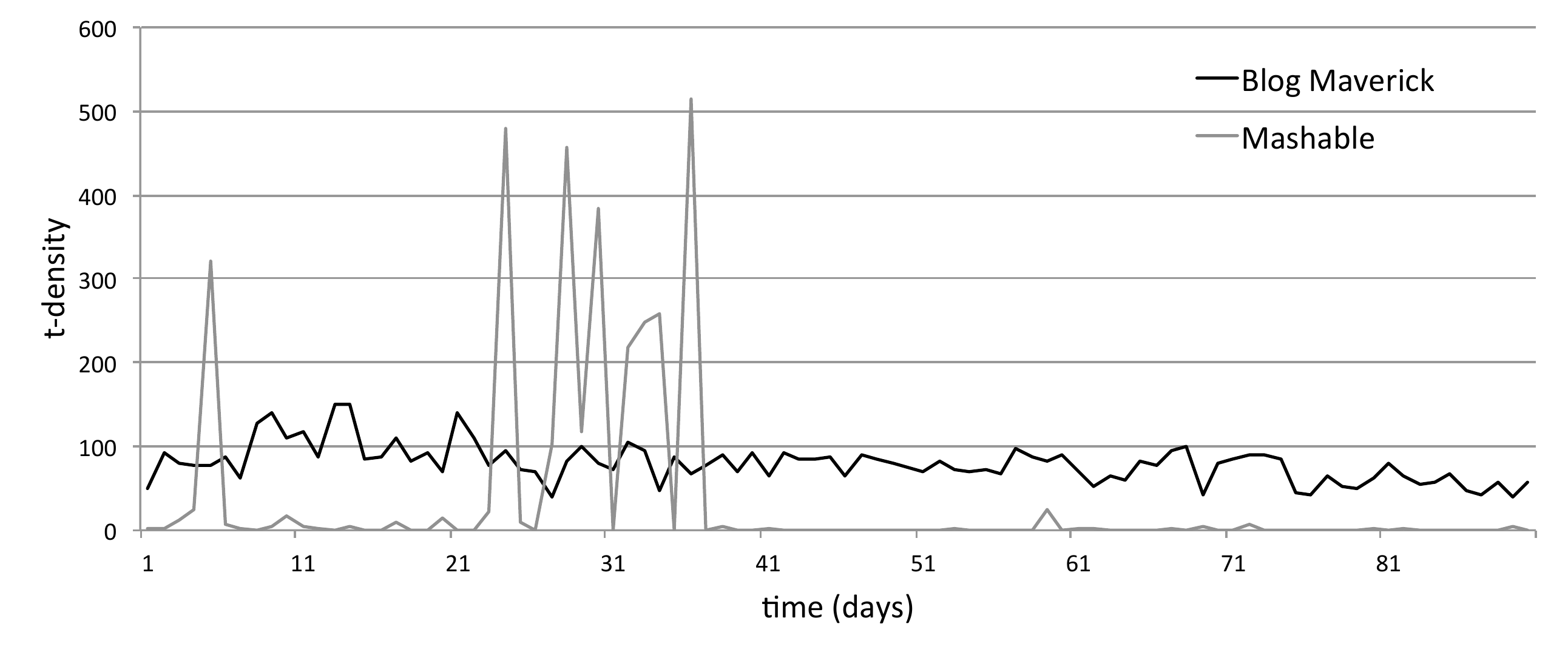} 
   \caption{Timeline of t-density (tweet per link) of two sources.}
   \label{fig:mashable}
\end{figure}

We plot the timeline of t-densities for a few sources and find that t-density of a source can vary greatly over time. Figure \ref{fig:mashable}  shows the t-density values belonging to the technology blog \textit{Mashable} and \textit{Blog Maverick}, a weblog of prominent entrepreneur, Mark Cuban. The t-density scores corresponding to each of these sources are 74 and 178 respectively. However, one can see that \textit{Mashable} has a more consistent t-density compared to \textit{Blog Maverick}.

\begin{figure}[ht]
\centering 
\subfloat[tweets and links]{\includegraphics[width=3.2in, trim = 0.5cm 0cm 0cm 0cm]{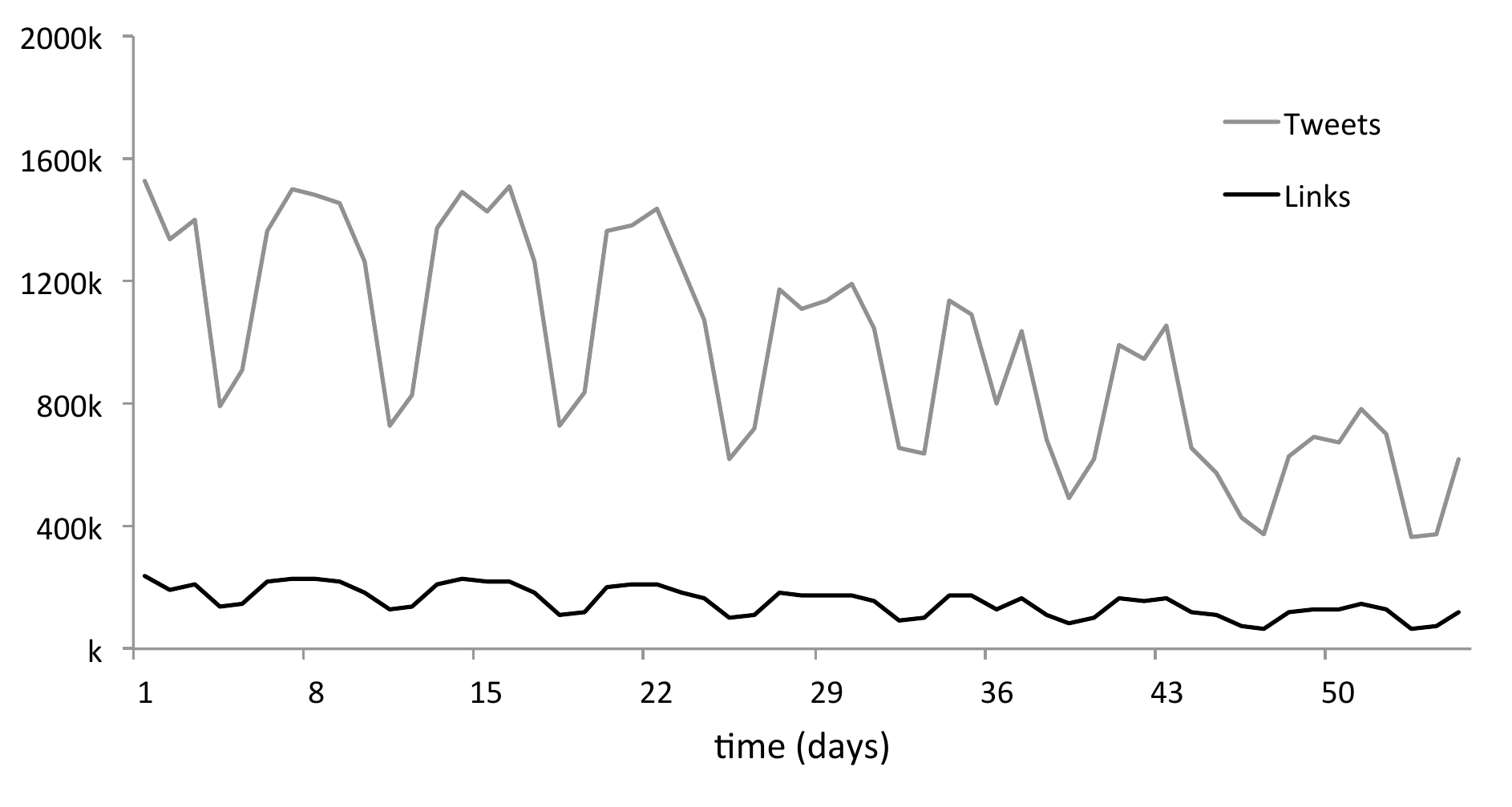}}\\
\subfloat[t-density]{\includegraphics[width=3.2in, trim = 1cm 0cm 0cm 0cm]{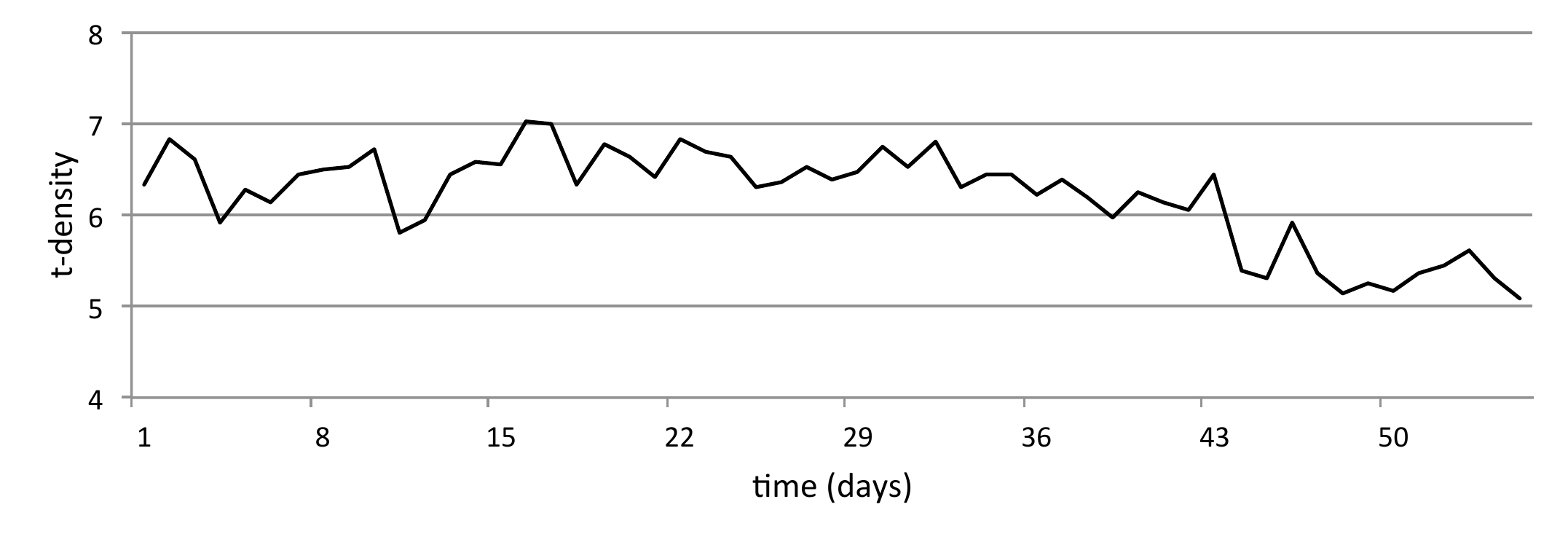}}
\caption{Temporal variations of tweets, links, and t-density over all sources}
\label{fig:temporal}
\end{figure}

In order to improve the score to reflect consistency we devise two methods; the first method is to smooth the measurements for each source by passing them through a low-pass filter. Second is to weight the score by the percentage of times a source's t-density is above the mean t-density over all sources, penalizing sources that drop low too often. The mean value of t-densities over all sources is 6.4. Figure \ref{fig:temporal} shows the temporal variations of tweets and links over all sources. Notice that while both tweets and links have a weekly cycle, the t-density (tweets over links) does not have this periodic nature.

\subsubsection{Are top traditional news sources the most propagated?}

As we assign scores to sources in our dataset, we are interested to know whether sources successful in this dataset are those that are conventionally considered prominent. 

Google News \footnote{http://news.google.com/} is one of the major aggregators and providers of news on the web. While inclusion in Google news results is free, Google uses its own criteria to rank the content and place some articles on its homepage, giving them more exposure. Freshness, diversity, and rich textual content are listed as the factors used by Google News to automatically rank each article as it is published. Because Google does not provide overall rankings for news sources, to get a rating of sources we use NewsKnife \footnote{http://www.newsknife.com}. NewsKnife is a service that rates top news sites and journalists based on analysis of article's positions on the Google news homepage and sub-pages internationally. We would like to know whether the sources that are featured more often on Google news (and thus deemed more prominent by Google and rated more highy by NewsKnife) are also those that become most popular on our dataset.

\begin{table}[htb]
\begin{center}
\begin{tabular}{|l|l|l|l|} 
\hline
&Total Links & Total Tweets & t-density\\
\hline
Correlation&0.57&0.35&-0.05\\
\hline
\end{tabular}
\caption{Correlation values between NewsKnife source scores and their performance on twitter dataset.}
\label{table:newsknifeCor}
\vspace{-0.25cm}
\end{center}
\end{table}

Accordingly we measure the correlation values for the 90 top NewsKnife sources that are also present in our dataset. The values are shown in Table \ref{table:newsknifeCor}. It can be observed that the ratings correlate positively with the number of links published by a source (and thus the sum of their tweets), but have no correlation (-0.05) with t-density which reflects the number of tweets that each of their links receives. For our source scoring scheme this correlation was about 0.7. 

Table \ref{table:popular} shows a list of top sources according to NewsKnife, as well as those most popular sources in our dataset. While NewsKnife rates more traditionally prominent news agencies such as Reuters and the Wall Street Journal higher, in our dataset the top ten sources (with highest t-densities) include sites such as Mashable, AllFacebook (the unofficial facebook blog), the Google blog, marketing blogs, as well as weblogs of well-known people such as Seth Godin's weblog and Mark Cuban's blog (BlogMaverick). It is also worth noting that there is a bias toward news and opinion on web marketing, indicating that these sites actively use their own techniques to increase their visibility on Twitter.

While traditional sources publish many articles, those more successful on the social web garner more tweets. A comparison shows that a NewsKnife top source such as The Christian Science Monitor received an average of 16 tweets in our dataset with several of its articles not getting any tweets. On the other hand, Mashable gained an average of nearly 1000 tweets with its least popular article still receiving 360 tweets. Highly ranked news blogs such as The Huffington Post perform relatively well in Twitter, possibly due to their active twitter accounts which share any article published on the site.

\begin{table}[h]
\begin{center}
\begin{tabular}{  | l | p{5cm} |}
\hline
NewsKnife & {\small \textit{Reuters, Los Angeles Times,  New York Times, Wall Street Journal, USA Today, Washington Post, ABC News, Bloomberg, Christian Science Monitor, BBC News} }\\
\hline
Twitter Dataset &{\small \textit{Blog Maverick, Search Engine Land, Duct-tape Marketing, Seth's Blog, Google Blog, Allfacebook, Mashable, Search Engine Watch}}\\
\hline
\end{tabular}
\caption{Highly rated sources on NewsKnife versus those popular on the Twitter dataset}
\label{table:popular}
\vspace{-0.25cm}
\end{center}
\end{table}

\section{Prediction}
In this work, we evaluate the performance of both regression and classification methods to this problem.
First, we apply regression to produce exact values of tweet counts, evaluating the results by the R-squared measure. Next we define popularity classes and predict which class a given article will belong to. The following two sections describe these methods and their results.  

 \begin{table}[h]
\begin{center}
  \begin{tabular}{p{2cm}p{5.5cm}}
    \hline
   Variable & Description\\ 
    \hline
	\(S\) & Source t-density score\\ 
	\(C\) & Category t-density score\\ 
	\(Subj\)& Subjectivity (0 or 1)\\ 
	\(Ent_{ct}\) & Number of named entities\\
	\(Ent_{max}\) & Highest score among named entities\\ 
	\(Ent_{avg}\) & Average score of named entities\\ 
    \hline
  \end{tabular}
  \caption{Feature set (prediction inputs)}
        \label{table:features}
\end{center}
\end{table}

\subsection{Regression}

Once score assignment is complete, each point in the data (i.e. a given news article) will correspond  to a point in the feature space defined by its category, subjectivity, named entity, and source scores. As described in the previous section, category, source, and named entity scores take real values while the subjectivity score takes a binary value of 0 or 1. Table \ref{table:features} lists the features used as inputs of regression algorithms. We apply three different regression algorithms - linear regression, k-nearest neighbors (KNN) regression and support vector machine (SVM) regression.

\begin{table}[h]
\begin{center}
  \begin{tabular}{lll}
    \hline
     &  Linear Regression &   SVM  Regression \\
         \hline  
    All Data &  0.34 &  0.32     \\
   Tech Category &   0.43  &   0.36  \\
   Within Twitter  &0.33&0.25\\
    \hline  
  \end{tabular}
  \caption{Regression Results}
    \label{table:regression}
\end{center}
\end{table}


Since the number of tweets per article has a long-tail distribution (as discussed previously in Figure \ref{fig:powerLaw100}), we performed a logarithmic transformation on the number of tweets prior to carrying out the regression. We also used the log of source and category scores to normalize these scores further. Based on this transformation, we reached the following relationship between the final number of tweets and feature scores.
\[ln(T) = 1.24ln(S) + 0.45ln(C)+0.1Ent_{max}-3\]
where \(S\) is the source t-density score, \(C\) is the category t-density score, and \(Ent_{max}\) is the maximum t-density of all entities found in the article. Equivalently,
\[T = S^{1.24} C^{0.45}e^{-(0.1Ent_{max}+3)}\]
with coefficient of determination \(R^{2}=0.258\). Note that the  \(R^{2}\) is the coefficient of determination and relates to the mean squared error and variance: \[R^{2} = 1 - {{MSE}\over{VAR}}\]

Alternatively, the following model provided improved results:
\[T^{0.45} = \left (0.2S-0.1Ent_{ct}- 0.1Ent_{avg}+0.2Ent_{max}\right )^{2} \]
with an improved \(R^{2}=0.34\). Using support vector machine (SVM) regression \cite{CC01a}, we reached similar values for \(R^{2}\) as listed in Table \ref{table:regression}. 


In K-Nearest Neighbor Regression, we predict the tweets of a given article using values from it's nearest neighbors. We measure the Euclidean distance between two articles based on their position in the feature space \cite{hastie2008elements}. Parameter $K$ specifies the number of nearest neighbors to be considered for a given article. Results with $K=7$ and $K=3$ for a 10k test set are R-sq= 0.05, with mean squared error of 5101.695. We observe that KNN performs increasingly more poorly as the dataset becomes larger.

\subsubsection{Category-specific prediction}
One of the weakest predictors in regression was the Category score. One of the reasons for this is that there seems to be a lot of overlap across categories. For example, one would expect \emph{World News} and \emph{Top News} to have some overlap, or the category \emph{USA} would feature articles that overlap with others as well. So the categories provided by Feedzilla are not necessarily disjoint and this is the reason we observe a low prediction accuracy.

To evaluate this hypothesis, we repeated the prediction algorithm for particular categories of content. Using only the articles in the Technology category, we reached an \(R^{2}\) value of 0.43, indicating that when employing regression we can predict the popularity of articles within one category  (i.e. Technology) with better results.

\subsection{Classification}

Feature scores derived from historical data on Twitter are based on articles that have been tweeted and not those articles which do not make it to Twitter. As discussed in Section \ref{sec:Data} this is evident in how the zero-tweet articles do not follow the linear trend of the rest of datapoints in Figure \ref{fig:powerLaw100}. Consequently, we do not include a zero-tweet class in our classification scheme and perform the classification by only considering those articles that were posted on twitter.  

Table \ref{table:classes} shows three popularity classes A (1 to 20 tweets), B (20 to 100 tweets), C (more than 100) and the number of articles in each class in the set of 10,000 articles. Table \ref{table:classification} lists the results of support vector machine (SVM) classification, decision tree, and bagging \cite{Hall:2009:WDM:1656274.1656278} for classifying the articles. All methods were performed with 10-fold cross-validation. We can see that classification can perform with an overall accuracy of 84\% in determining whether an article will belong to a low-tweet, medium-tweet, or high-tweet class. 

In order to determine which features play a more significant role in prediction, we repeat SVM classification leaving one of the features out at each step. We found that publication source plays a more important role compared to other predictors, while subjectivity, categories, and named entities do not provide much improvement in prediction of news popularity on Twitter. 

\subsubsection{Predicting Zero-tweet Articles}
We perform binary classification to predict which articles will be at all mentioned on Twitter (zero tweet versus nonzero tweet articles). Using SVM classification we can predict --with 66\% accuracy-- whether an article will be linked to on twitter or whether it will receive zero tweets. We repeat this operation by leaving out one feature at a time to see a change in accuracy. We find that the most significant feature is the source, followed by its category. Named entities and subjectivity did not provide more information for this prediction. So despite one might expect, we find that readers overall favor neither subjectivity nor objectivity of language in a news article.

It is interesting to note that while category score does not contribute in prediction of popularity within Twitter, it does help us determine whether an article will be at all mentioned on this social network or not. This could be due to a large bias toward sharing technology-related articles on Twitter.

\begin{table}[h]
\begin{center}
  \begin{tabular}{lll}
    \hline
      Class name&  Range of tweets  &  Number of articles\\
    \hline
A&1--20 & 7,600\\
B&20--100 &1,800\\
C&100--2400 &600\\
    \hline
  \end{tabular}
  \caption{Article Classes}    
  \label{table:classes}
\end{center}
\end{table}

\begin{table}[h]
\begin{center}
  \begin{tabular}{p{5cm}p{2cm}}
    \hline
Method&  Accuracy  \\
\hline
Bagging& 83.96\%\\
J48 Decision Trees & 83.75\%\\
SVM & 81.54\% \\
Naive Bayes&77.79\%\\
    \hline  
  \end{tabular}
  \caption{Classification Results}
        \label{table:classification}
\end{center}
\end{table}

\section{Discussion and Conclusion}

In this work we predicted the popularity of news items on Twitter using features extracted from the content of news articles. We have taken into account four features that cover the spectrum of the information that can be gleaned from the content - the source of the article, the category, subjectivity in the language and the named entities mentioned. Our results show that while these features may not be sufficient to predict the exact number of tweets that an article will garner, they can be effective in providing a range of popularity for the article on Twitter. We achieved an overall accuracy of 84\% using classifiers. It is important to bear in mind that while it is intriguing to pay attention to the most popular articles --those that become viral on the web-- a great number of articles spread in medium numbers. These medium levels can target highly interested and informed readers and thus the mid-ranges of popularity should not be dismissed. 

Interestingly we have found that in terms of number of retweets, the top news sources on twitter are not necessarily the conventionally popular news agencies and various technology blogs such as Mashable and the Google Blog are very widely shared in social media. Overall, we discovered that one of the most important predictors of popularity was the source of the article. This is in agreement with the intuition that readers are likely to be influenced by the news source that disseminates the article. On the other hand, the category feature did not perform well. One reason for this is that we are relying on categories provided by Feedzilla, many of which overlap in content. Thus a future task is to extract categories independently and ensure little overlap. Combining other layers of complexity described in the introduction opens up the possibility of better prediction. It would be interesting to incorporate network factors such as the influence of individual propagators to this work.

\bibliographystyle{aaai}

\begin{thebibliography}{}

\bibitem[\protect\citeauthoryear{Agarwal \bgroup et al.\egroup
  }{2008}]{Agarwal:2008:IIB:1341531.1341559}
Agarwal, N.; Liu, H.; Tang, L.; and Yu, P.~S.
\newblock 2008.
\newblock Identifying the influential bloggers in a community.
\newblock In {\em Proceedings of the international conference on Web search and
  web data mining}, WSDM '08,  207--218.
\newblock New York, NY, USA: ACM.

\bibitem[\protect\citeauthoryear{Alias-i.}{2008}]{lingpipe}
Alias-i.
\newblock 2008.
\newblock Lingpipe 4.1.0.
\newblock http://alias-i.com/lingpipe.

\bibitem[\protect\citeauthoryear{Chang and Lin}{2011}]{CC01a}
Chang, C.-C., and Lin, C.-J.
\newblock 2011.
\newblock {LIBSVM}: A library for support vector machines.
\newblock {\em ACM Transactions on Intelligent Systems and Technology}
  2:27:1--27:27.
\newblock Software available at http://www.csie.ntu.edu.tw/~cjlin/libsvm.

\bibitem[\protect\citeauthoryear{Cosley \bgroup et al.\egroup
  }{2010}]{ICWSM101530}
Cosley, D.; Huttenlocher, D.; Kleinberg, J.; Lan, X.; and Suri, S.
\newblock 2010.
\newblock Sequential influence models in social networks.
\newblock In {\em 4th International Conference on Weblogs and Social Media}.

\bibitem[\protect\citeauthoryear{Gruhl \bgroup et al.\egroup
  }{2004}]{DBLP:journals/sigkdd/GruhlLGT04}
Gruhl, D.; Liben-Nowell, D.; Guha, R.~V.; and Tomkins, A.
\newblock 2004.
\newblock Information diffusion through blogspace.
\newblock {\em SIGKDD Explorations} 6(2):43--52.

\bibitem[\protect\citeauthoryear{Hall \bgroup et al.\egroup
  }{2009}]{Hall:2009:WDM:1656274.1656278}
Hall, M.; Frank, E.; Holmes, G.; Pfahringer, B.; Reutemann, P.; and Witten,
  I.~H.
\newblock 2009.
\newblock The weka data mining software: an update.
\newblock {\em SIGKDD Explor. Newsl.} 11:10--18.

\bibitem[\protect\citeauthoryear{Hastie, Tibshirani, and
  Friedman}{2008}]{hastie2008elements}
Hastie, T.; Tibshirani, R.; and Friedman, J.
\newblock 2008.
\newblock {\em The elements of statistical learning: data mining, inference,
  and prediction}.
\newblock Springer series in statistics. Springer.

\bibitem[\protect\citeauthoryear{Kempe, Kleinberg, and
  Tardos}{2003}]{DBLP:conf/kdd/KempeKT03}
Kempe, D.; Kleinberg, J.~M.; and Tardos, {\'E}.
\newblock 2003.
\newblock Maximizing the spread of influence through a social network.
\newblock In {\em KDD},  137--146.
\newblock ACM.

\bibitem[\protect\citeauthoryear{Kim, Kim, and Cho}{2011}]{6036808}
Kim, S.-D.; Kim, S.-H.; and Cho, H.-G.
\newblock 2011.
\newblock Predicting the virtual temperature of web-blog articles as a
  measurement tool for online popularity.
\newblock In {\em IEEE 11th International Conference on Computer and
  Information Technology (CIT)},  449 --454.

\bibitem[\protect\citeauthoryear{Lee, Moon, and
  Salamatian}{2010}]{DBLP:conf/webi/LeeMS10}
Lee, J.~G.; Moon, S.; and Salamatian, K.
\newblock 2010.
\newblock An approach to model and predict the popularity of online contents
  with explanatory factors.
\newblock In {\em Web Intelligence},  623--630.
\newblock IEEE.

\bibitem[\protect\citeauthoryear{Lerman and
  Ghosh}{2010}]{DBLP:conf/icwsm/LermanG10}
Lerman, K., and Ghosh, R.
\newblock 2010.
\newblock Information contagion: An empirical study of the spread of news on
  digg and twitter social networks.
\newblock In {\em ICWSM}.
\newblock The AAAI Press.

\bibitem[\protect\citeauthoryear{Lerman and
  Hogg}{2010}]{DBLP:conf/www/LermanH10}
Lerman, K., and Hogg, T.
\newblock 2010.
\newblock Using a model of social dynamics to predict popularity of news.
\newblock In {\em WWW},  621--630.
\newblock ACM.

\bibitem[\protect\citeauthoryear{Leskovec, Adamic, and
  Huberman}{2007}]{DBLP:journals/tweb/LeskovecAH07}
Leskovec, J.; Adamic, L.~A.; and Huberman, B.~A.
\newblock 2007.
\newblock The dynamics of viral marketing.
\newblock {\em TWEB} 1(1).

\bibitem[\protect\citeauthoryear{Leskovec, Backstrom, and
  Kleinberg}{2009}]{DBLP:conf/kdd/LeskovecBK09}
Leskovec, J.; Backstrom, L.; and Kleinberg, J.~M.
\newblock 2009.
\newblock Meme-tracking and the dynamics of the news cycle.
\newblock In {\em KDD},  497--506.
\newblock ACM.

\bibitem[\protect\citeauthoryear{Leskovec \bgroup et al.\egroup
  }{2007}]{Leskovec07cascadingbehavior}
Leskovec, J.; Mcglohon, M.; Faloutsos, C.; Glance, N.; and Hurst, M.
\newblock 2007.
\newblock Cascading behavior in large blog graphs.
\newblock In {\em In SDM}.

\bibitem[\protect\citeauthoryear{Rogers}{1995}]{rogers1995diffusion}
Rogers, E.
\newblock 1995.
\newblock {\em {Diffusion of innovations}}.
\newblock Free Pr.

\bibitem[\protect\citeauthoryear{Romero, Meeder, and
  Kleinberg}{2011}]{Romero2011}
Romero, D.~M.; Meeder, B.; and Kleinberg, J.
\newblock 2011.
\newblock Differences in the mechanics of information diffusion across topics:
  idioms, political hashtags, and complex contagion on twitter.
\newblock In {\em Proceedings of the 20th international conference on World
  wide web}, WWW '11,  695--704.
\newblock New York, NY, USA: ACM.

\bibitem[\protect\citeauthoryear{Salganik, Dodds, and
  Watts}{2006}]{salganik2010}
Salganik, M.~J.; Dodds, P.~S.; and Watts, D.~J.
\newblock 2006.
\newblock Experimental study of inequality and unpredictability in an
  artificial cultural market.
\newblock {\em Science} 311(5762):854--856.

\bibitem[\protect\citeauthoryear{Simkin and Roychowdhury}{2008}]{Simkin2008}
Simkin, M.~V., and Roychowdhury, V.~P.
\newblock 2008.
\newblock {A theory of web traffic}.
\newblock {\em EPL (Europhysics Letters)} 82(2):28006.

\bibitem[\protect\citeauthoryear{Sun \bgroup et al.\egroup
  }{2009}]{DBLP:conf/icwsm/SunRML09}
Sun, E.; Rosenn, I.; Marlow, C.; and Lento, T.~M.
\newblock 2009.
\newblock Gesundheit! modeling contagion through facebook news feed.
\newblock In {\em ICWSM}.
\newblock The AAAI Press.

\bibitem[\protect\citeauthoryear{Szab{\'o} and
  Huberman}{2010}]{DBLP:journals/cacm/SzaboH10}
Szab{\'o}, G., and Huberman, B.~A.
\newblock 2010.
\newblock Predicting the popularity of online content.
\newblock {\em Commun. ACM} 53(8):80--88.

\bibitem[\protect\citeauthoryear{Tatar \bgroup et al.\egroup
  }{2011}]{Tatar2011}
Tatar, A.; Leguay, J.; Antoniadis, P.; Limbourg, A.; de~Amorim, M.~D.; and
  Fdida, S.
\newblock 2011.
\newblock Predicting the popularity of online articles based on user comments.
\newblock In {\em Proceedings of the International Conference on Web
  Intelligence, Mining and Semantics}, WIMS '11,  67:1--67:8.
\newblock New York, NY, USA: ACM.

\bibitem[\protect\citeauthoryear{V\'azquez \bgroup et al.\egroup
  }{2006}]{PhysRevE.73.036127}
V\'azquez, A.; Oliveira, J.~a.~G.; Dezs\"o, Z.; Goh, K.-I.; Kondor, I.; and
  Barab\'asi, A.-L.
\newblock 2006.
\newblock Modeling bursts and heavy tails in human dynamics.
\newblock {\em Phys. Rev. E} 73:036127.

\bibitem[\protect\citeauthoryear{Wu and Huberman}{2007}]{Wu06112007}
Wu, F., and Huberman, B.~A.
\newblock 2007.
\newblock Novelty and collective attention.
\newblock {\em Proceedings of the National Academy of Sciences}
  104(45):17599--17601.

\bibitem[\protect\citeauthoryear{Yang and
  Leskovec}{2011}]{DBLP:conf/wsdm/YangL11}
Yang, J., and Leskovec, J.
\newblock 2011.
\newblock Patterns of temporal variation in online media.
\newblock In {\em WSDM},  177--186.
\newblock ACM.

\bibitem[\protect\citeauthoryear{Yu, Chen, and
  Kwok}{2011}]{DBLP:conf/sbp/YuCK11}
Yu, B.; Chen, M.; and Kwok, L.
\newblock 2011.
\newblock Toward predicting popularity of social marketing messages.
\newblock In {\em SBP}, volume 6589 of {\em Lecture Notes in Computer Science},
   317--324.
\newblock Springer.

\bibitem[\protect\citeauthoryear{Zhou \bgroup et al.\egroup
  }{2010}]{Zhou:2010:IRT:1964858.1964875}
Zhou, Z.; Bandari, R.; Kong, J.; Qian, H.; and Roychowdhury, V.
\newblock 2010.
\newblock Information resonance on twitter: watching iran.
\newblock In {\em Proceedings of the First Workshop on Social Media Analytics},
  SOMA '10,  123--131.
\newblock New York, NY, USA: ACM.

\end{thebibliography}

\end{document}